\documentclass[useAMS,usenatbib]{mn2e}
\usepackage{graphicx}

\title[Constraints on quintessence and new physics from fundamental constants]{Constraints on quintessence and new physics from fundamental constants}
\author[Rodger I. Thompson]{Rodger I. Thompson$^{1}$\thanks{E-mail:
rit@email.arizona.edu (RIT)}\\
$^{1}$Steward Observatory, University of Arizona, Tucson, AZ 85721, USA}
\begin{document}

\date{Accepted xxxx. Received xxxx; in original form xxxx}

\pagerange{\pageref{firstpage}--\pageref{lastpage}} \pubyear{2011}

\maketitle

\label{firstpage}

\begin{abstract}
Changes in the values of the fundamental constants $\mu$, the proton to electron
mass ratio, and $\alpha$, the fine structure constant due to rolling scalar fields 
have been discussed both in the context of cosmology and in new physics such as 
Super Symmetry (SUSY) models.  This article examines the changes in these fundamental
constants in a particular example of such fields, freezing and thawing slow roll 
quintessence.  Constraints are placed on the product of a cosmological quantity, w,
the equation of state parameter, 
and the square of the coupling constants for $\mu$ and $\alpha$ with the field, 
$\zeta_{x}$, ($x = \mu,\alpha$) using the existing observational limits on the 
values of $\frac{\Delta x}{x}$.  Various examples of slow rolling quintessence
models are used to further quantify the constraints.  Some of the examples
appear to be rejected by the existing data which strongly suggests that conformation
to the values of the fundamental constants in the early universe is a standard test 
that should be applied to any cosmological model or suggested new physics.
\end{abstract}


\section{Introduction} \label{s-intro}

The values of the fundamental constants such as the fine structure constant
$\alpha$ and the proton to electron mass ratio $\mu$ in the early universe
provide important constraints and tests for cosmological theories such as
quintessence and new physics models that suggest a coupling between the 
constants and rolling scalar fields. It was pointed out more than 35 years 
ago \citep{thm75} that changes in the value of the fundamental constant $\mu$
produce changes in molecular spectra such that observations of molecular
spectra in high redshift objects can track the value of $\mu$ in the
early universe.  The recent advent of large telescopes with sensitive high
resolution spectrometers and, most importantly, very accurate measurements
of the wavelengths of the molecular hydrogen Lyman and Werner band transitions
\citep{uba07,mal10} have made such observations possible.  Most of the 
relevant observations have been of the molecular hydrogen Lyman and Werner 
electronic absorption transitions in high redshift Damped Lyman Alpha systems
(DLAs) \citep{kin09,wen08,thm09,mal10,kin11}.  These observations have restricted
the change in $\mu$ to $\frac{\Delta \mu}{\mu} \le 10^{-5}$ at redshifts up
to 3.  Radio observations have established limits on $\frac{\Delta \mu}{\mu}$
on the order of $10^{-6}$ using a comparison between the
inversion transition of ammonia and the rotational transitions of other 
molecules at a redshift of 0.685 \citep{mur08} and at a
redshift of 0.89 \citep{mul11}.  At redshifts between 2 and 4 \citet{cur11}
compared the frequencies of CO rotational lines to the frequency of the fine 
structure transition of neutral carbon.  This comparison, however, does not
measure $\mu$ directly but rather $F \equiv \frac{\alpha^2}{\mu}$.  It is 
interesting, therefore, to
see what constraints or limits these observations can put on cosmological 
models and the necessity of new physics not consistent with the standard 
model.

The situation regarding $\alpha$ is less clear given the conflicting claims on
the variation \citep{mur04} or constancy \citep{cha04} of $\alpha$ with even 
claims of a spatial dipole variation of $\alpha$ \citep{web11}. Given this state 
of uncertainty we will concentrate mainly on the limits imposed by the 
constancy of $\mu$ but will also consider the consequences of a variation in
$\alpha$ at the end of the analysis.

\section[]{Cosmological Model} \label{s-quin}

As a definite example of a cosmological model with a potential $V(\phi)$ defined by 
a rolling scalar field $\phi$ we will examine slow rolling freezing and thawing quintessence 
models following the discussion of \citet{sch08} and \citet{dut11}, hereinafter DS. In this case the 
dark energy is due to a minimally coupled scalar field $\phi$ governed by the equation
of motion
\begin{equation}
\ddot{\Phi} + 3H \dot{\phi} + \frac{dV}{d\phi} = 0
\end{equation}
where H is the Hubble parameter in units where $8 \pi G = 1$.  The slow roll 
conditions are given by
\begin{equation}\label{eq-sr1}
\lambda^2 \equiv (\frac{1}{V} \frac{dV}{d\phi})^2 \ll 1
\end{equation}
and
\begin{equation}
|\frac{1}{V} \frac{d^2V}{d\phi^2}| \ll 1
\end{equation}
which leads to a very flat potential. Thawing solutions are where the equation of
state parameter $w = p_{de}/\rho_{de}$ is initially near -1 and evolves away from -1
while freezing solutions start with $w$ not equal to -1 and evolve toward -1 at the
present day. (DS) have shown that these cosmologies follow trajectories
given by
\begin{equation} \label{eq-qw}
1 + w = \frac{1}{3} \lambda_0^2[\frac{1}{\sqrt{\Omega_{\phi}}}-(\frac{1}{\Omega_{\phi}}-1)
(\tanh^{-1}(\sqrt{\Omega_{\phi}}) + C)]^2
\end{equation} 
where C is determined by an early condition on $w$ and $\Omega_{\phi}$
\begin{equation}
C = \pm \frac{\sqrt{3(1+w_i)}\Omega_{\phi_i}}{\lambda_0}
\end{equation}
and the dark energy density factor $\Omega_{\phi}$ is given by
\begin{equation}\label{eq-omega}
\Omega_{\phi} = [1+(\Omega_{\phi 0}^{-1} - 1)a^{-3}]^{-1}
\end{equation}
In (\ref{eq-omega}) $a$ is the scale factor of the universe normalized to 1 at
the present day and the subscript $0$ refers to the present epoch.  Thawing 
solutions are given by the special case where $C=0$.
The parameter $\lambda$ is necessarily small as given by equation \ref{eq-sr1}
and is considered to be constant and equal to $\lambda_0$ at all times.

\begin{figure}
  \vspace*{75pt}
\resizebox{\textwidth}{!}{\includegraphics[0in,0in][11in,1.3in]{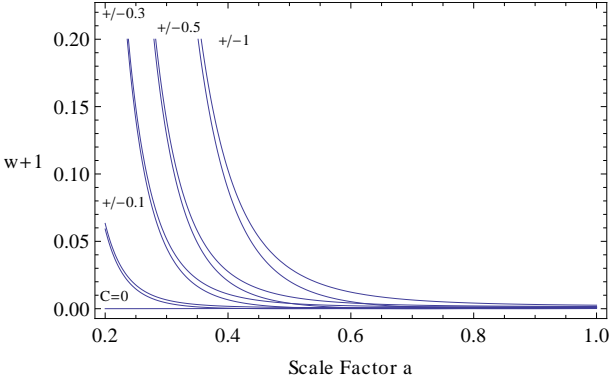}}
  \caption{The evolution of w+1 as a function of the scale factor a for various
values of C. 
The scale factor a is normalized to 1 at the present epoch.  The track
for C=0 is the thawing solution which starts near w+1=0 and ends with a value slightly
larger than 0 at the present epoch. The values of $\lambda_0$ and $\Omega_{\phi i}$
are given in the last paragraph of section \ref{s-quin}. Note that values of $1 + w$
greater than 0.05 are beyond the range considered by DS.} \label{fig-wc}
\end{figure}

(DS) consider several cases with values of C set to -1, -0.5, -0.1, 0.,
0.1, 0.5, and 1 with $w_i = -0.95$, $\lambda_0 = -0.08$, considered constant by DS,
 and $\Omega_{\phi i}$ adjusted to yield the listed values of C.  We will use this 
suite of cases plus the added cases of C = -0.3 and C = 0.3 to
determine what constraints are imposed by the observed values of $\mu$. 
Figure~\ref{fig-wc} shows the evolution of w+1 as a function of the scale factor
a for the various values of C. Due to their large excursions from $w + 1 = 0$
at high z the larger absolute values of C probably violate the slow roll condition.
That is why we have added the C =+/-0.3 cases.   All of the curves are for
freezing quintessence except of the curve for C=0 which is for thawing
quintessence.  Although hard to see at the scale of Figure~\ref{fig-wc}, the C=0 
case for thawing quintessence has $w+1 \neq 0$ at $a=1$.

\section{New Physics and $\mu$} \label{s-np}

Since $\mu$ does not vary with time in the standard model any variation of $\mu$
must include new physics.  This discussion follows the methodology described in
\citet{nun04}, hereinafter NL which presumes a linear, non-varying, coupling of $\mu$ to the 
rolling scalar field.  There can be many variations of the coupling between $\mu$
and the scalar field but this represents one of the simplest models.  (NL)
actually consider the variation of the fine structure constant $\alpha$, however,
most new physics models assume that $\alpha$ and $\mu$ vary in the same manner
and are connected in the following manner
\begin{equation}  \label{eq-amu}
\frac{\dot{\mu}}{\mu} \sim \frac{\dot{\Lambda}_{QCD}}{\Lambda_{QCD}} -\frac{\dot{\nu}}
{\nu} \sim R \frac{\dot{\alpha}}{\alpha}
\end{equation}
eg. \citet{ave06}. In (\ref{eq-amu}) $\Lambda_{QCD}$ is the QCD scale, $\nu$ is 
the Higgs vacuum expectation value and R is often considered to be on the order
of -40 to -50 \citep{ave06} but it is highly indeterminate.  The variation in $\alpha$
and consequentially $\mu$ is given as
\begin{equation} \label{eq-dmu}
\frac{\Delta \mu}{\mu} = R \zeta_{\alpha} \kappa (\phi - \phi_0) = \zeta_{\mu} \kappa (\phi - \phi_0)
\end{equation}
where $\zeta_x$ ($x = \alpha, \mu$) is the coupling constant, 
$\kappa = \frac{\sqrt{8 \pi}}{m_p}$ and $m_p$ is the Planck mass. Based on the work of
\citet{cop04} (NL)
place limits on $|\zeta_{\alpha}|$ of $|\zeta_{\alpha}| \sim 10^{-4} - 10^{-7}$ which
translate to limits on $|\zeta_{\mu}|$ of $|\zeta_{\mu}| \sim 4 \times 10^{-3} -4 \times 
10^{-6}$ for $R = -40$.

The equation 
of state is given in terms of the dark energy pressure $p_{\phi}$, dark energy
density $\rho_{\phi}$ and dark energy potential $V(\phi)$ by
\begin{equation} \label{eq-w}
w \equiv \frac{p_{\phi}}{\rho_{\phi}} = \frac{\dot{\phi}^2 - 2 V(\phi)}{\dot{\phi}^2 + 2 V(\phi)}
\end{equation}
(NL) then show that $w +1$ is also given by
\begin{equation} \label{eq-wa}
w + 1 = \frac{(\kappa \phi')^2}{3 \Omega_{\phi}}
\end{equation}
where $\Omega_{\phi}$ is the dark matter energy density. Here $\dot{\phi}$ and $\phi'$ indicate 
differentiation with respect to cosmic time and to $N = \log{a}$ respectively where $a$ is 
again the scale factor. It follows from (\ref{eq-dmu}) that
\begin{equation} \label{eq-phip}
\phi' = \frac{\mu'}{\kappa \zeta_{\mu} \mu}
\end{equation}
Substitution of (\ref{eq-phip}) into (\ref{eq-wa}) then links $w$ and
$\mu'$ as
\begin{equation} \label{eq-wmu}
w +1 = \frac{(\mu'/\mu)^2}{3 \zeta_{\mu}^2 \Omega_{\phi}} = \frac{(\alpha'/\alpha)^2}{3 \zeta_{\alpha}^2 \Omega_{\phi}}
\end{equation}
which links the evolution of $w$, $\mu$ and $\alpha$.

\section{Relating $\mu$, $\alpha$ and $w$}

\begin{figure}
  \vspace*{75pt}
\resizebox{\textwidth}{!}{\includegraphics[0in,0in][11in,1.3in]{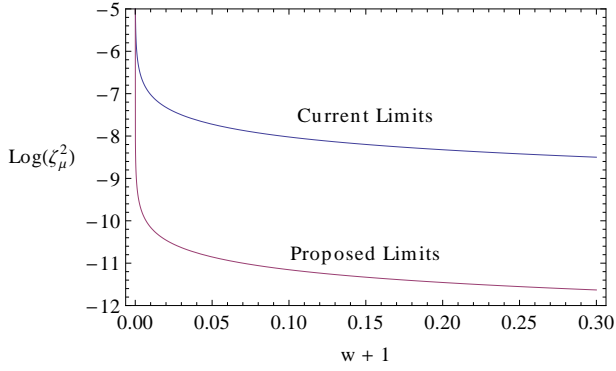}}
  \caption{The parameter space in terms of $\log (\zeta_{\mu})^2$ and $(w+1)$ defined by the
current limit on $\Delta \mu / \mu$ and for one that is 50 times lower than the present limit. 
All of the space above the two curves is forbidden in terms of the models and parameters 
presented here.} \label{fig-wze}
\end{figure}

Now that we have two different equations for $w +1$, one involving only cosmological
factors, the equation of state parameter, in the quintessence model discussed in section 
\ref{s-quin} and one involving
new physics in section \ref{s-np} we can see what constraints the observations place
on the parameters for these models. From~(\ref{eq-wmu}) we see that limits
on $\Delta x / x$ place limits on the product $\zeta_{x}^2(w+1)$ if we assume
that the cosmology of $\Omega_{\phi}$ is known from equation ~\ref{eq-omega}. Note
that the use of equation ~\ref{eq-omega} restricts the results to the slow roll conditions. 
Using the current limit on $\Delta \mu / \mu \leq 10^{-5}$ we can define regions of
parameter space in a $(\zeta_{\mu})^2$ - $(w+1)$ landscape that are forbidden or allowed
by those limits. Figure~\ref{fig-wze} shows that space for the current limit on 
$\Delta \mu / \mu$ and for one where the limit is 50 times more stringent, a possibility
using observations from expected new spectrometers such as PEPSI on the LBT.

It is obvious that as $w+1$ approaches 0 that the constraint of $\zeta_{\mu}^2$
diminishes rapidly.  This has to be so since it is the coupling with $\phi$ that
both drives $w+1$ away from 0 and produces the change in $\mu$.  If the improved
limits are achieved then there will be a more than 1000 times stronger constraint 
on $\zeta_{\mu}^2$ even for very small values of $w+1$.

\begin{figure}
   \vspace*{75pt}
\resizebox{\textwidth}{!}{\includegraphics[0in,0in][11in,1.3in]{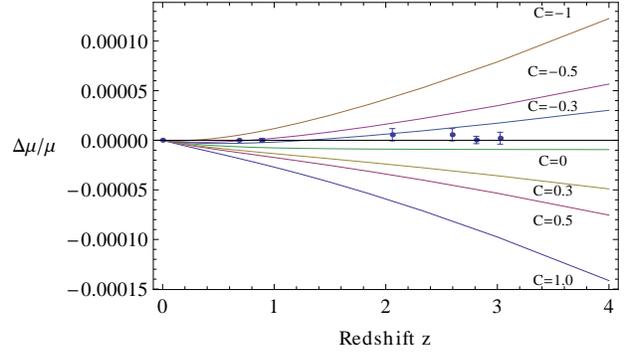}}
  \caption{Plot of $\Delta \mu / \mu$ as a function of redshift $z = \frac{1}{a} - 1$. 
The $C$ values are labeled
in the figure and $\zeta{\mu}$ is set to $-4 \times 10^{-4}$.  All of the solutions are for freezing
quintessence except for $C = 0$ which is the thawing solution.  The data points are taken
from the information in Table~\ref{tab-obs} below along with the stated errors. The plotted points for
objects with more than one reference are the most restrictive constraints. The erroneous \citet{rei06}
and \citet{uba07} results are not shown.} \label{fig-dmuz}
\end{figure}

In the following we will use $\mu$ as a definite example but the result for $\alpha$
is exactly the same with $\zeta_{\alpha}$ replacing $\zeta_{\mu}$.
Setting the right hand sides of equations ~\ref{eq-wmu} and ~\ref{eq-qw} equal to each
other we get
\begin{equation}
\frac{(\mu'/\mu)^2}{3 \zeta_{\mu}^2 \Omega_{\phi}} = \frac{1}{3} \lambda_0^2\{\Omega_{\phi}^{-1/2}-(\Omega_{\phi}^{-1}-1)(\tanh^{-1}(\Omega_{\phi}^{1/2}) + C)\}^2
\end{equation}
which yields
\begin{equation}
\frac{\mu'}{\mu} = \zeta_{\mu} \lambda_0 \Omega_{\phi}^{1/2}\{\Omega_{\phi}^{-1/2}-(\Omega_{\phi}^{-1}-1)(\tanh^{-1}(\Omega_{\phi}^{1/2}) + C)\}
\end{equation}
We then use that $\mu' = a(\frac{d \mu}{d a})$ to produce
\begin{eqnarray}
\frac{d\mu}{\mu} = \zeta_{\mu} \lambda_0 \int_{1}^{a} \Omega_{\phi}^{1/2} \{\Omega_{\phi}^{-1/2}-
(\Omega_{\phi}^{-1}-1) \nonumber \\
\times (\tanh^{-1}(\Omega_{\phi}^{1/2}) + C)\}a^{-1}da
\end{eqnarray}
\begin{eqnarray}
\frac{d\mu}{\mu} = \zeta_{\mu} \lambda_0 \int_{1}^{a}\{1-(\Omega_{\phi}^{-1/2} - \Omega_{\phi}^{1/2}) 
  \nonumber \\
\times (\tanh^{-1} (\Omega_{\phi}^{1/2})+C)\}a^{-1}da
\end{eqnarray}
We next substitute in~(\ref{eq-omega}) for $\Omega_{\phi}$ to get 
\begin{eqnarray} \label{eq-ddmu}
\frac{d\mu}{\mu} = \zeta_{\mu} \lambda_0 \int_{1}^{a} \{1-[(1+(\Omega_{0}^{-1}-1)a^{-3})^{-1/2} \nonumber \\
-(1+(\Omega_{0}^{-1}-1)a^{-3})^{1/2}] \nonumber \\
\times [\tanh^{-1} (1+(\Omega_{0}^{-1}-1)a^{-3})^{1/2}+C]\}a^{-1}da
\end{eqnarray}
Since we know that $\mu$ is constant to one part in $10^5$ to observable limits we can treat $\mu$ in the denominator of the left side of~(\ref{eq-ddmu}) as a constant and finally get
\begin{eqnarray} \label{eq-deltamu}
\frac{\Delta \mu}{\mu} = \zeta_{\mu} \lambda_0 \int_{1}^{a} \{1-[(1+(\Omega_{0}^{-1}-1)a^{-3})^{-1/2} \nonumber \\
-(1+(\Omega_{0}^{-1}-1)a^{-3})^{1/2}] \nonumber \\
\times [\tanh^{-1} (1+(\Omega_{0}^{-1}-1)a^{-3})^{1/2}+C]\}a^{-1}da
\end{eqnarray}
Equation~\ref{eq-deltamu} can be numerically integrated using Mathematica\footnote{Copyright 
1988-2011 Wolfram Research Inc.} to show the evolution of the value of $\mu$ as a function
of the scale factor a normalized to 1 at the present time and $\Omega_0$ set to 0.7.  (DS) use
a constant $\lambda$ of $\lambda_0= -0.08$, satisfying the first slow roll condition and consider
a range of the constant $C$ between -1 and 1.  Negative values of C correspond to a 
field rolling down the potential and positive values indicate the field initially rolling up 
the potential. (DS) indicate that the latter case is unlikely but they consider it
for completeness.  Figure~\ref{fig-dmuz} shows the results of the integration plotted as a 
function of redshift for a value of $\zeta_{\mu} =-4 \times 10^{-4}$ which is consistent
with $\zeta_{\alpha} = 10^{-5}$ and $R=-40$.. It is evident from~(\ref{eq-deltamu}) 
that the results scale linearly with $\zeta_{\mu}$.

A comparison of the observed value of $\Delta \mu / \mu$ and the $\Delta \mu / \mu$ values 
predicted by the models with $\zeta_{\mu}$ set to $-4.0 \times 10^{-4}$ at a redshift of 
3 is given in the first 7 entries of Table~\ref{tab-preobs}.  It is evident from both the 
table and Figure~\ref{fig-dmuz} that most of the quintessence models
considered in this paper, except for the thawing model, are ruled out by the observations if
the coupling constant $\zeta_{\mu}$ is as high as $-4 \times 10^{-4}$ and the magnitude
of $C$ is significantly greater than 0.  Since the value of $C$ sets the initial value
of $w+1$ in the freezing models the small value of $\Delta \mu / \mu$ favors models
where $w+1$ is small or 0 in the early universe. It is certainly 
possible to bring any of the quintessence models into compliance by lowering the value
of either $\zeta_{\mu}$ or equivalently R in~(\ref{eq-amu}). However, these calculations show that
using parameters that are common in the literature yields results that are incompatible
with the present limits on $\Delta \mu / \mu$.  It therefore suggests that the limits on
the variance of fundamental constants is a rigorous test that should be applied to any proposed
cosmological models or new physics.  The slow roll conditions and the linear coupling
of the fundamental constants with the field are conditions that provide a minimal
change in the constants, therefore, these results impose an even stronger constraint
on quintessence type models with much steeper potentials.

\section{Imposing a condition on $\alpha$}
\begin{figure}
   \vspace*{75pt}
\resizebox{\textwidth}{!}{\includegraphics[0in,0in][11in,1.3in]{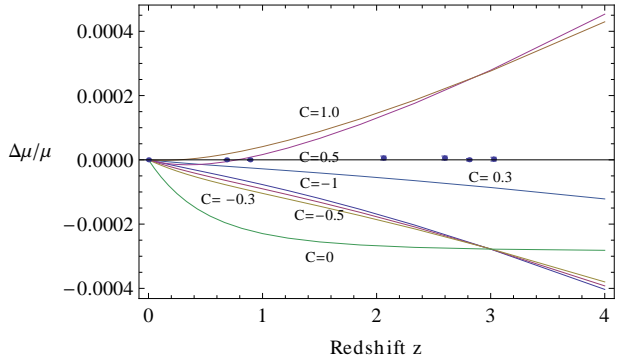}}
  \caption{Plot of $\Delta \mu / \mu$ as a function of redshift with $\zeta_{\alpha}$
adjusted to give $\alpha$ of $\Delta \alpha /\alpha = 7 \times 10^{-6}$ at a redshift
of 3. The $C$ values are labeled in the plot. Note that at z=1 the order of the plots 
is that C=-0.3 is the lowest and C=-1 is the highest of the three closely bunched solutions. 
The data points are the same as in figure~\ref{fig-dmuz}.} \label{fig-da}
\end{figure}

There have been persistent claims in the literature that the value of $\alpha$ was different
in the past than it is now.  If we use the most recent claim of an altered previous
value of $\alpha$ of $\Delta \alpha /\alpha = 7 \times 10^{-6}$ at redshifts around
3 \citep{web11} we can check what conditions this imposes on $\Delta \mu / \mu$. In 
Figure~\ref{fig-da} we have adjusted $\zeta_{\alpha}$ to give a value of 
$\Delta \alpha/\alpha= 7 \times 10^{-6}$ for each value of C with R set to -40.

It is clear from Figure~\ref{fig-da} that none of the curves fit the high redshift
$\Delta \mu / \mu$ measurements.  Table~\ref{tab-preobs} indicates that at a 
redshift of three the predictions differ from the observations by about a factor 
of 130. The thawing model ($C=0$) is a particularly bad fit even at low redshifts
 since $\zeta_{\alpha}$ has to be set to a very high value to achieve the claimed 
value of $\Delta \alpha /\alpha$. For this model the low redshift radio results of 
\citet{mur08} are more than a factor of 1000 less than the predicted value. 
These results significantly increase the tension between the $\alpha$ and $\mu$ 
observations unless the magnitude of R in~(\ref{eq-amu}) is about -0.3 at
an epoch when the redshift is about 3. This value of R is very different from
the expected value of around -40 derived from generic GUT models where the strong
coupling constant and the Higgs Vacuum Expectation Value run exponentially faster
than $\alpha$ \citep{ave06}.  In this context either the expectations from Super
Symmetry considerations are flawed or the conclusions from the observations
indicating a change in the value of $\alpha$ in both space and time are in error.

\begin{table}
 \begin{minipage}{80mm}
\begin{tabular}{ccc}
\hline
$\frac{(\Delta \mu/\mu)_{0bs}}{(\Delta \mu/\mu)_{Mod}}$ &$\zeta_{\mu}$  & C \\

\hline
-0.022 & $-4.0 \times 10^{-4}$ & -1.0 \\
-0.039 & $-4.0 \times 10^{-4}$ & -0.5 \\
-0.059 & $-4.0 \times 10^{-4}$ & -0.3 \\
-0.23 & $-4.0 \times 10^{-4}$& 0.0 \\
-0.12 & $-4.0 \times 10^{-4}$ & 0.3 \\
0.060 & $-4.0 \times 10^{-4}$ & 0.5 \\
0.026 & $-4.0 \times 10^{-4}$ & 1.0 \\
-0.0076 & $-1.14 \times 10^{-3}$ & -1.0 \\
-0.0076 & $-2,08 \times 10^{-3}$ & -0.5 \\
-0.0076 & $-3.1 \times 10^{-3}$ & -0.3 \\
-0.0076 & $-1.2 \times 10^{-2}$ & 0.0 \\
0.0076 & $-6.5 \times 10^{-4}$ & 0.3 \\
0.0075 & $-3.2 \times 10^{-3}$ & 0.5 \\
0.0076 & $-1.4 \times 10^{-3}$ & 1.0 \\
\hline
\end{tabular}
\caption{Model Predictions versus Observations at Z=3. The first 7 entries
are for $\zeta_{\mu}$ held at $-4.0 \times 10^{-4}$. In the last 7 entries
the value of $\zeta_{\mu}$ is adjusted to produce a $\Delta \alpha /\alpha$
value of $7 \times 10^{-6}$ with R held at -40.} \label{tab-preobs}
\end{minipage}
\end{table}

\section*{Conclusions}

Constraints on the values of the fundamental constants $\mu$ and $\alpha$ now impose
meaningful constraints on both cosmological models and new physics in terms of the
product of the equation of state parameter $(w+1)$ and a new physics term $\zeta_x^2$ ($x=\mu,
\alpha$) as shown in figure~\ref{fig-wze}.  Given a specific cosmological model such 
as slow roll quintessence and expected parameters for the coupling constants $\zeta_x$
predictions of the changes in the fundamental constants $\Delta x/x$ can be made.
These predictions, using parameters common in the literature, are discrepant with
the observational constraints at high redshift by up to two orders of magnitude for 
some cases. This argues for the inclusion of the values of the fundamental constants
in the early universe into the commonly accepted tests for either cosmological models
or new physics.  It is certainly true that the new physics and cosmological parameters
can be adjusted to satisfy the fundamental constant constraints.  The main point, however,
is that the parameters are no longer completely free but must be contained within 
observational bounds.  Improvements in the measurement of the fundamental constants
can significantly improve the constraints and provide even more stringent bounds
on new physics and cosmology. An improvement by a factor of 50 on the current constraint 
on $\Delta \mu/\mu$ of $\leq 10^{-5}$ produces more than a thousand fold improvement
of the constraint on $(\zeta_{\mu})^2(w+1)$.  Such an improvement should be possible
with new instrumentation such as the PEPSI spectrometer on the Large Binocular Telescope.

\section*{Acknowledgments}
The author would like to acknowledge very helpful discussions with C.J.A.P. Martins,
F. Ozel, D. Psaltis, D. Marrone and. J. Bechtold as well as the very useful comments
of an anonymous referee.

\appendix

\section{Current determinations of $\Delta \mu/\mu$}
\label{s-curs}
Table~\ref{tab-obs} lists the current determinations of $\Delta \mu/\mu$ in distant
galaxies and in the Milky Way. A subset of the most recent constraints are used in
figures~\ref{fig-dmuz} and~\ref{fig-da}.  In particular the erroneous positive
results of \citet{rei06} have not been included.

\begin{table}
 \begin{minipage}{80mm}
\begin{tabular}{llll}
\hline
Object & Reference & Redshift & $\Delta\mu/\mu$ \\

\hline
Q0347-383 & \citet{wen08} & 3.0249 & $(2.1 \pm 6) \times 10^{-6}$\\
Q0347-383 & \citet{kin09} & 3.0249 & $(8.2 \pm 7.4) \times 10^{-6}$\\
Q0347-383 & \citet{thm09} & 3.0249 &$(-2.8 \pm 1.6) \times 10^{-5}$\\
Q0528-250 & \citet{kin09} & 2.811 & $(1.4 \pm 3.9)\times 10^{-6}$ \\
Q0528-250 & \citet{kin11} & 2.811 & $(0.3 \pm 3.7)\times 10^{-6}$ \\
Q0405-443 & \citet{thm09} & 2.5974 & $(3.7 \pm 14) \times 10^{-6}$ \\
Q0405-443 & \citet{kin09} & 2.5974 & $(10.1 \pm 6.2) \times 10^{-6}$ \\
J2123-005 & \citet{mal10} & 2.059  & $5.6 \pm 6.2) \times 10^{-6}$\\
PKS 1830-211 &\citet{mul11} & 0.89 & $\le 2. \times 10^{-6}$ \\
B0218+357 & \citet{mur08} & 0.6847 & $\le 0.18 \times 10^{-6}$ \\
Milky Way & \citet{mol09} & 0.0 & $(4 - 14) \times 10^{-8}$ \\
\hline
\end{tabular}
\caption{Recent Astronomical $\Delta\mu/\mu$ Measurements}  \label{tab-obs}
\end{minipage}
\end{table}
\label{lastpage}
\end{document}